\documentclass[12pt]{article}
\pdfoutput=1
\usepackage{graphicx,amssymb,amsfonts,amsthm,amsmath,amssymb,amscd,amstext, color}
\def\exp{\operatorname{exp}}

\def\be{\begin{equation}}
\def\ee{\end{equation}}

\begin{document}

\thispagestyle{empty}

\begin{flushright}
YITP-SB-15-46
\end{flushright}

\begin{center}
\vspace{1cm} { \LARGE {\bf
Relativistic Hydrodynamics and Non-Equilibrium Steady States
}}
\vspace{1.1cm}

Michael Spillane and Christopher P.~Herzog\\
\vspace{0.8cm}
{ \it C.~N.~Yang Institute for Theoretical Physics \\
Department of Physics and Astronomy \\
Stony Brook University, Stony Brook, NY  11794}

\vspace{0.8cm}

\end{center}

\begin{abstract}
\noindent

We review recent interest in the relativistic Riemann problem as a method for generating a non-equilibrium steady state.  In the version of the problem under consideration, the initial conditions consist of a planar interface between two halves of a system held at different temperatures in a hydrodynamic regime.  The new double shock solutions are in contrast with older solutions that involve one shock and one rarefaction wave.  We use numerical simulations to show that the older solutions are preferred.  Briefly we discuss the effects of a conserved charge.  Finally, we discuss deforming the relativistic equations with a nonlinear term and how that deformation affects the temperature and velocity in the region connecting the asymptotic fluids.

\end{abstract}

\pagebreak
\setcounter{page}{1}

\section{Introduction}

The Riemann problem in hydrodynamics is an initial value problem where two equilibrium fluids are joined by a discontinuity.  The solution for the case of relativistic fluids was first solved in 1948 \cite{Taub}.  It has since been studied in other papers including \cite{Thompson, Mach}.  The key feature of these solutions is a rarefaction (adiabatic) region joined to a constant temperature region and then a shock discontinuity.\footnote{%
 For a discussion of these issues in the context of the quark-gluon plasma, see for example ref.\ \cite{Rischke:1995ir}.
}

The problem has seen renewed interest recently as an example of a steady state system which is not in thermal equilibrium (NESS).  This type of NESS was first studied in 1+1 dimensional CFT's \cite{Bernard:2012je} and later extended to hydrodynamical descriptions of higher dimensional CFT's \cite{Bhaseen2013, Chang}.  Finally, it was considered for a CFT deformed by a relevant operator \cite{Pourhasan}.  The papers \cite{Bhaseen2013,Pourhasan}, however, miss the important rarefaction region and as a result their values for the NESS temperature as well as the rate of growth of the NESS are incorrect.\footnote{%
The paper \cite{Chang} restricts to a regime where the temperature difference is very small, and the difference between
the two shock solution and the solution involving a rarefaction region correspondingly minuscule.
}

In this paper we review both the older solution with the rarefaction region as well as the more recent solution which we call the two shock solution.  Using a numerical simulation of relativistic hydrodynamics we show that the rarefaction solution matches the numerical simulation better than the two shock solution.  We also consider a CFT deformed by a relevant operator.  We calculate a phase diagram for the temperature of the NESS. The phase diagram we present differs from the two shock solution.

\section{Ideal Hydrodynamics}
We begin with the stress tensor of a perfect 
 fluid and conserved currents
\begin{align}
T^{\mu\nu} &= (e+p) u^\mu u^\nu+p \eta^{\mu\nu}, \\
J_{i}^\mu &= q_i u^\mu.
\end{align}
We mean perfect in the sense of having no dissipation.  Here $e$ is the energy density, $q_i$ a charge density, and $p$ the pressure.
We have introduced a four velocity $u^\mu$ such that $u^2  = u^\mu u^\nu \eta_{\mu\nu} = -1$.  At rest the fluid is described by $u^\mu = (1, \vec 0)$.  We work in mostly plus signature with the Minkowski metric tensor $\eta^{\mu\nu} = (-,+, \ldots, +)$.  
The conservation equations of energy and momentum are given by
\begin{align}
\label{EOM}&\partial_\mu T^{\mu\nu} = 0,\\
&\partial_\mu J_{i}^\mu = 0.
\end{align}
The conservation equations are combined with an equation of state $e= e(p)$.  In this paper we will largely focus on a linear equation of state 
\begin{align}
p = c_s^2 e  , 
\end{align}
 where $c_s$ is the speed of sound.  One important example for us is a conformal fluid in $d$ spatial dimensions, where $c_s^2 = 1/d$.

We are interested in flows depending only on a single spatial variable, arbitrarily chosen to be $x$, and time.  We will perform a change of variables to the local fluid velocity $v_i= u^i/u^t$.  In these variables the stress tensor conservation equations become

\begin{align}
\label{con1}\partial_t\left[(e+p) \gamma^2 -p\right]+\partial_x\left[(e+p) \gamma^2 v_x\right] =0, \\
\label{con2}
\partial_t\left[(e+p) \gamma^2 v_x\right]+\partial_x\left[(e+p) \gamma^2 v_x^2+p\right] =0, \\
\label{con3}\partial_t\left[(e+p) \gamma^2 \vec v_T \right]+\partial_x\left[(e+p) \gamma^2 v_x \vec v_T\right] =0.
\end{align}
We have introduced  $\gamma = 1/\sqrt{1-v_x^2-v_T^2}$ and also $\vec v_T$, the fluid velocity in the spatial directions perpendicular to $x$. 

When there is a shock discontinuity we use the Rankine-Hugoniot jump condition to determine the relationship between conserved quantities on alternate sides of the shock \cite{Pourhasan,evans_pde}.  For a conservation law of the form 
\begin{align}
\partial_t Q(t,x) +\partial_x F(t,x) = 0,\\
\label{conservation} u_s [Q] = [F], \\
[Q] = Q_L-Q_R, \\
[F] = F_L-F_R,
\end{align}
where $Q_L$ ($Q_R$) and $F_L$ ($F_R$) is the value of $Q$ and $F$ to the left (right) of the shock and $u_s$ is the velocity of the shock.  For the case of a perfect fluid, the jump conditions are given by
\begin{align}
\label{conservation1}u_s [T^{tt}] = [T^{tx}],\\
\label{conservation2}u_s [T^{xt}] = [T^{xx}], \\
\label{conservation3}u_s[J_i^t] = [J_i^x] .
\end{align}
Equations like (\ref{conservation3}) can exhibit what are known as contact discontinuities.  These have a jump in the conserved quantities. However, there is no transportation of particles across the discontinuity.  In the case of (\ref{conservation3}), such a contact discontinuity can occur when $u_x/u_t = u_s$. The jump condition is trivially satisfied and the change in the conserved quantity can be arbitrary across the shock.  

\section{Double Shock solution}

We are interested in solving the Riemann problem, where two semi-infinite fluids of different temperatures are brought into contact.  An interesting feature of the resulting fluid flow is a non equilibrium steady state (NESS) that forms in the expanding region between the two semi-infinite fluids.  Recently two papers \cite{Bhaseen2013, Pourhasan} have presented a solution that is not completely correct.  Let us quickly review their work to see where the problem occurs.   

They start with the EOS $p = c_s^2 e$.  The initial conditions are that of two systems (with energies $e_L$ and $e_R$) brought into thermal contact.  They assume that two shocks propagate away from each other, leaving the NESS in between.  The central region is assumed to have a constant fluid velocity $v$.  
(We will suppress the $x$ subscript on the fluid velocity $v$ in what follows and set $v_T=0$.) 
The conservation laws (\ref{conservation1})-(\ref{conservation3}) imply
\begin{align}
u_L = -c_s \sqrt{\frac{c_s^2\chi +1}{\chi+c_s^2}}, \quad u_R =  c_s \sqrt{\frac{\chi+c_s^2}{c_s^2\chi +1}}, \\
e = \sqrt{e_L e_R} \ , \quad \frac{v}{c_s} = \frac{\chi-1}{\sqrt{(c_s^2 \chi +1)(\chi+c_s^2)}},
\end{align}
where $\chi = \sqrt{e_L/e_R}$.  The velocity of the left and right moving shocks are $u_L$ and $u_R$ respectively.  

This solution however is invalid because it violates the Entropy Condition \cite{evans_pde}.  
This condition is most easily stated when the conservation conditions are written in characteristic form,
\begin{align}
\label{entropycond}\partial_t \vec{u}+B(\vec{u})\partial_x \vec{u}=0,
\end{align}
where here $\vec u(x,t) = (p(x,t), v(x,t))$.  
The eigenvalues $\lambda_\pm$ of $B$ are
\begin{align}
\lambda_\pm = \frac{v\pm c_s}{1\pm v c_s^2}.
\end{align}
These characteristics correspond to the local right and left moving speeds of sound at a given space-time point in the fluid.
(Reassuringly, $\lambda_\pm \to \pm c_s$ in the limit where the background fluid velocity vanishes, $v=0$.)

The entropy condition requires that for solutions involving a shock,
characteristics end on a shock discontinuity rather than begin on it.  By ending on the shock, information is lost and entropy should increase.  In contrast, in order for characteristics to begin on a shock, boundary conditions need to be specified, decreasing the entropy.  More precisely, consider a right moving shock, $u_s>0$.  Let $\lambda_R$ and $\lambda_L$ be eigenvalues of $B$ immediately to the right and left of the shock, respectively.  We should take the eigenvalues corresponding to right moving characteristics.  For the characteristics to end on the shock, it is necessary that
\begin{align}
\lambda_L>u_s>\lambda_R \ .
\end{align}
A similar condition is also required for a left moving shock, taking now the left moving characteristic eigenvalues instead.

Rewriting the entropy condition (\ref{entropycond}) for the right moving shock yields
\begin{align}
\frac{u_R^2-c_s^2 +(1-c_s^2 )c_s u_R}{c_s^2 (u_R^2- c_s^2) + u_R ( 1-c_s^2)} > u_R>c_s.
   \end{align}
This condition is true for $u_R \in (c_s,1),$ which is true for $\chi>1.$  Therefore a shock is a valid solution for the wave moving into the colder medium.  However for $u_R \in (c_s^2, c_s)$, which is true for $\chi< 1$, neither inequality holds.  Thus the entropy condition rules out a shock moving into the hotter region.  
(We could also analyze separately the left moving shock, but the physics is invariant under parity.)\footnote{%
 When $c_s = 1$, which holds for a conformal field theory in 1+1 dimensions, $u_R = \lambda_\pm = 1$ as well, and the entropy condition is satisfied (and saturated) for both the left and right moving shocks. We will see below that in this degenerate case, the two shock solution becomes identical with the solution involving a rarefaction wave.
}

While we know from this analysis that the double shock solution is unphysical, it turns out to be very close to the actual solution in some situations.  Given the simplicity of the double shock solution, it is interesting to consider adding a conserved charge.  This addition requires us to include a contact discontinuity.  A contact discontinuity is a discontinuity in one variable that travels at the local fluid velocity.  For such a discontinuity the Rankine-Hugoniot jump condition is trivially satisfied and the change in the variable can be arbitrary. In this case the discontinuity is in the conserved charge. The result is the splitting of the NESS region into two NESS with distinct charges but equal velocities and pressures.  The resulting charge densities are 
\begin{align}
q_1= \frac{q_L \sqrt{c_s^2  \chi +1}}{\sqrt{\chi } \sqrt{c_s^2 +\chi }},\\
q_2= \frac{q_R \sqrt{\chi } \sqrt{c_s^2 +\chi }}{\sqrt{c_s^2  \chi +1}}. 
\end{align}
where $q_1$ is the charge density in the region adjacent to $q_L$, $q_2$ is the charge density in the region adjacent to $q_R$.\footnote{The entropy condition is trivially satisfied for shock discontinuities.}

\begin{figure}[!htb]
\centering
  \centering
  \includegraphics[width=0.48\linewidth]{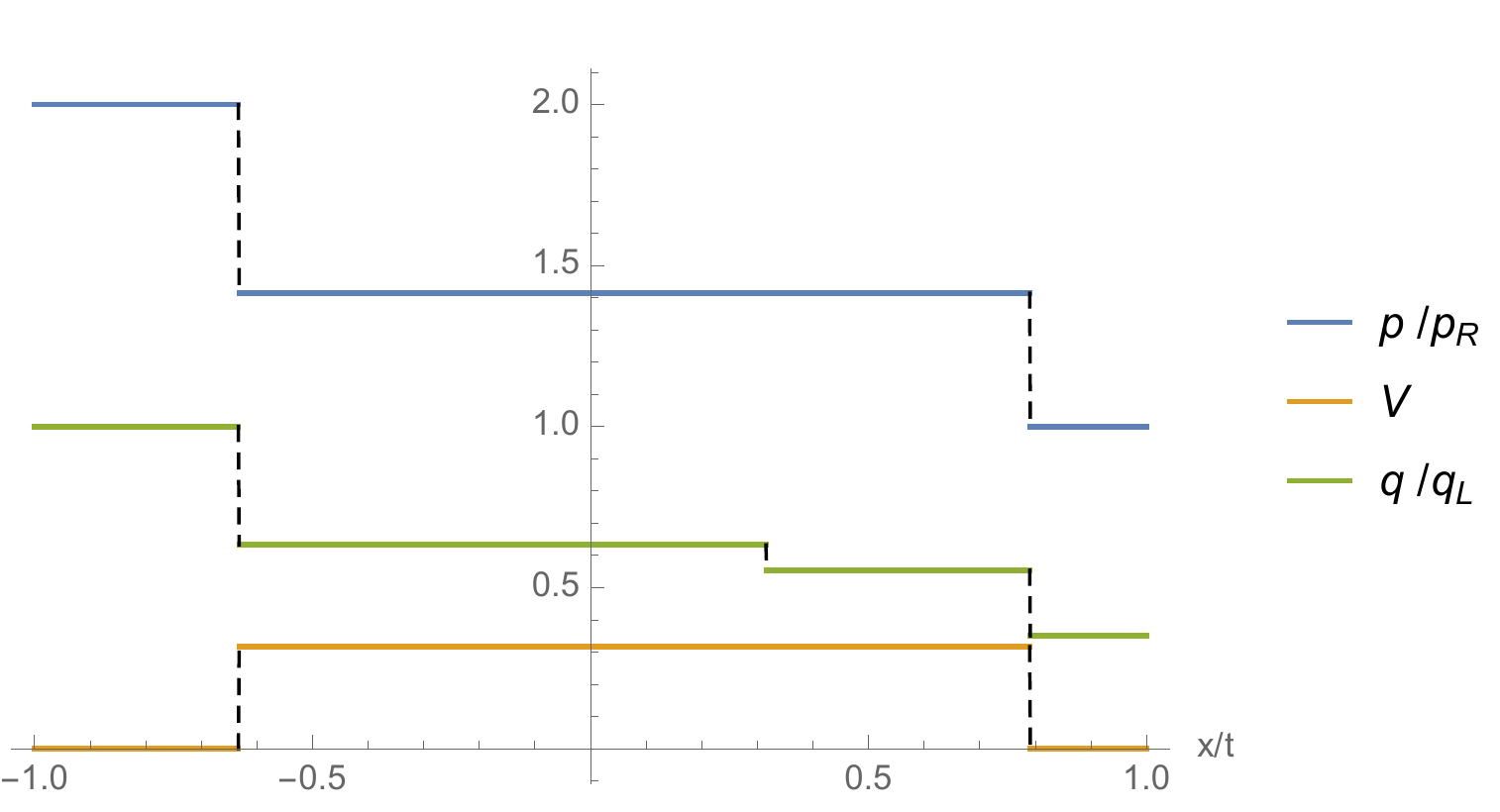}
  \quad
  \includegraphics[width=0.48\linewidth]{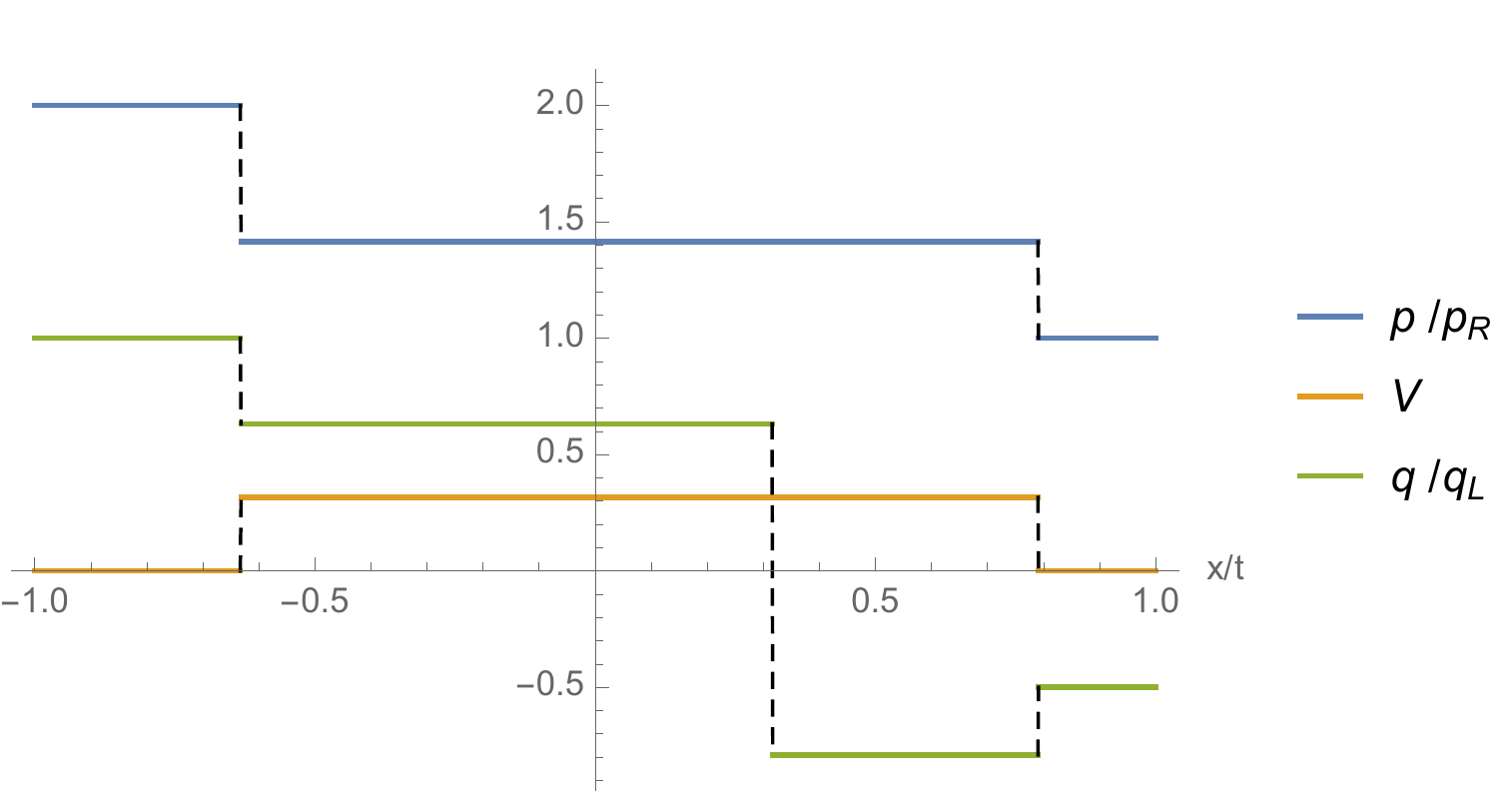}
 \caption{
 %\small 
 Plot of triple shock solutions.  The regions from left to right are $L$, 1, 2 and $R$.}

\end{figure}

\section{Adiabatic flow}

We now need to replace the shock that does not satisfy the entropy condition with a smooth solution.  The solution can be found in previous papers.  As can be concluded by considering characteristics, the shock solution should be replaced with a fan of characteristics \cite{Thompson,Mach,evans_pde}.  There is a characteristic for each value of the dimensionless ratio $\xi =x/t$.  Therefore we will search for a solution that depends only on $\xi =x/t$.  Such a solution would correspond to an adiabatic expansion.

It is simple to check that for $s =  p^{1/(1+c_s^2)}$, eqs.\ (\ref{con1}--\ref{con3}) imply
\begin{align}
\label{entropyeq}\partial_t(s \gamma)+\partial(s \gamma v_x) = 0.
\end{align}
After switching to coordinates $\xi = x/t$, we can combine equations (\ref{con3}) and (\ref{entropyeq}) and obtain
\begin{align}
(\xi - v_x)\frac{d}{d\xi}(p^{\kappa}\gamma v_T)=0 \ , 
\end{align}
where we have defined
\begin{align}
\kappa \equiv \frac{c_s^2}{1+c_s^2} \ .
\end{align}
The solution in the adiabatic region is then
\begin{align}
v_T &= \frac{\alpha  \sqrt{1-v_x(\xi )^2}}{\sqrt{\alpha ^2+p^{2 \kappa }}}  \ , \\
v_x &= \frac{\pm\left(\xi ^2-1\right) c_s p^{\kappa} \sqrt{\alpha
   ^2 (1-c_s^2 )+p^{2 \kappa}}+\xi  (c_s^2-1) \left(\alpha
   ^2+p^{2 \kappa}\right)}{\alpha ^2 (c_s^2 -1)+\left(\xi ^2 c_s^2
   -1\right) p^{2 \kappa}} \ ,   \\
p^{\kappa} &=\frac{\alpha ^2 (c_s^2 -1) \left(\frac{1-\xi }{\xi +1}\right)^{\mp\frac{c_s}{2}}}{4 c_1}+c_1 \left(\frac{1-\xi }{\xi
   +1}\right)^{\frac{\pm c_s}{2}} \ , \\
q &= \exp\left(\int d\xi \frac{\frac{\alpha ^2+p^{2 \kappa}}{1-\xi ^2}+\frac{\alpha ^2 \kappa
   \partial_x p}{p }+\frac{\alpha ^2 \kappa c_s  p^{\kappa-1} \partial_x p}{ \sqrt{p^{2 \kappa}-\alpha ^2 (c_s^2 -1)}}}{\alpha
   ^2+p^{2 \kappa}}\right)  \ .
\end{align} 
If we take the limit of zero tangential velocity ($\alpha = 0$), there are two solutions: 
\begin{align}
v_\pm(\xi) &= \frac{c_s \pm \xi}{c_s \xi \pm1} \ , \\
p_\pm(\xi) &= \overline p \left(\frac{1-\xi}{1+\xi}\right)^{\pm\frac{1+c_s^2}{2c_s}} \ , \\
q_\pm(\xi) &= \overline q \left(\frac{1-\xi}{1+\xi}\right)^{\pm\frac{1}{2c_s}} \ .
\end{align}

To solve the Riemann problem we need to match this adiabatic region onto a NESS region and a shock.  Without loss of generality we can choose $p_L> p_R.$ We then have a shock moving to the right at speed $u_s$.  In the left region, the disturbance moves at the speed of sound as can be seen by setting $v_\pm = 0$.  To the right of the shock $v = 0$ and $p = p_R$. To the left of the shock $v=V$ and $p=p_0$.  Then the jump conditions are
\begin{align}
p_0 (c_s^2  V (u_s V+1)+u_s-V)+p_R u_s
   \left(V^2-1\right) =0 \ , \\
 p_0 (c_s^2 +V (c_s^2  u_s+u_s-V))+p_R c_s^2
   \left(V^2-1\right)=0 \ .
\end{align}
Ideally when we put everything together we would get $V$ and $u_s$ as functions of $\chi$.  The best we were able to achieve were parametric expressions of $V$ and $\chi$ as functions $u_s$:
\begin{align}
V&= \frac{u_s^2  - c_s^2}{u_s(1-c_s^2)} \ , \\
\chi(u_s)^2 &= \frac{(u_s-c_s^2 ) (c_s^2 +u_s)}{c_s^2(1 -  u_s^2)}\left(\frac{1+u_s}{1-u_s}\right)^{\frac{c_s^2 +1}{2 c_s}}
   \left(\frac{u_s-c_s^2 }{c_s^2 +u_s}\right)^{\frac{c_s^2 +1}{2
  c_s}} \ ,  \\
   p_0 &= p_L \left(\frac{(1-u_s)(u_s+c_s^2)}{(1+u_s)(u_s-c_s^2)} \right)^{\frac{c_s^2 +1}{2
   c_s}} \ , 
\end{align}
where $\chi = \sqrt{p_L/p_R}$. 
We can then add in the charge which has a contact discontinuity.  
\begin{align}
q_1 &= q_L\left(\frac{(1-u_s)(u_s+c_s^2)}{(1+u_s)(u_s-c_s^2)} \right)^{\frac{1}{2
  c_s}} \ , \\
 q_2 &= \frac{q_R u_s \sqrt{u_s^2-c_s^4}}{c_s^2 \sqrt{1-u_s^2}}\ . 
\end{align}
\begin{figure}[!htb]
\centering
  \centering
  \includegraphics[width=0.48\linewidth]{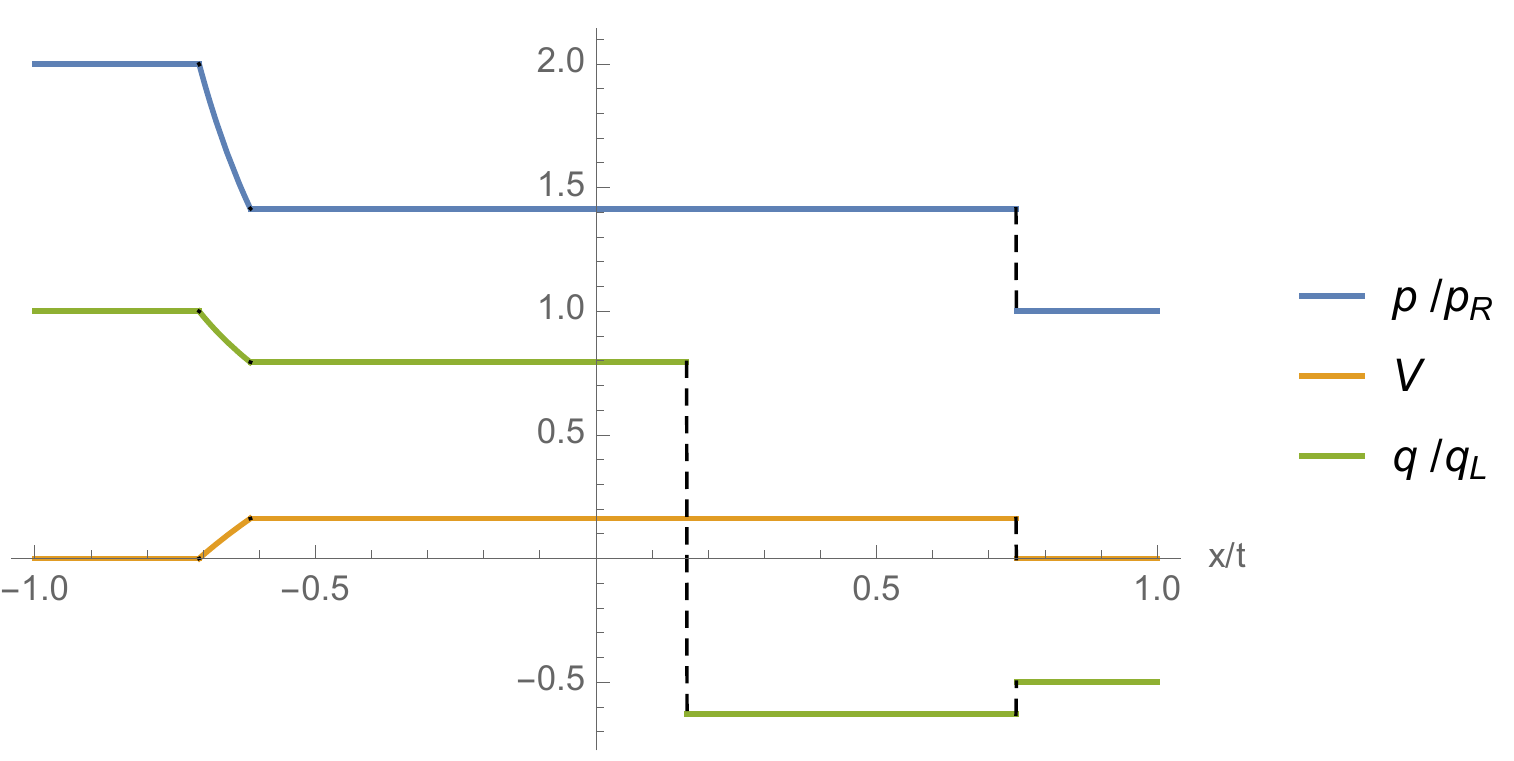}
  \quad
  \includegraphics[width=0.48\linewidth]{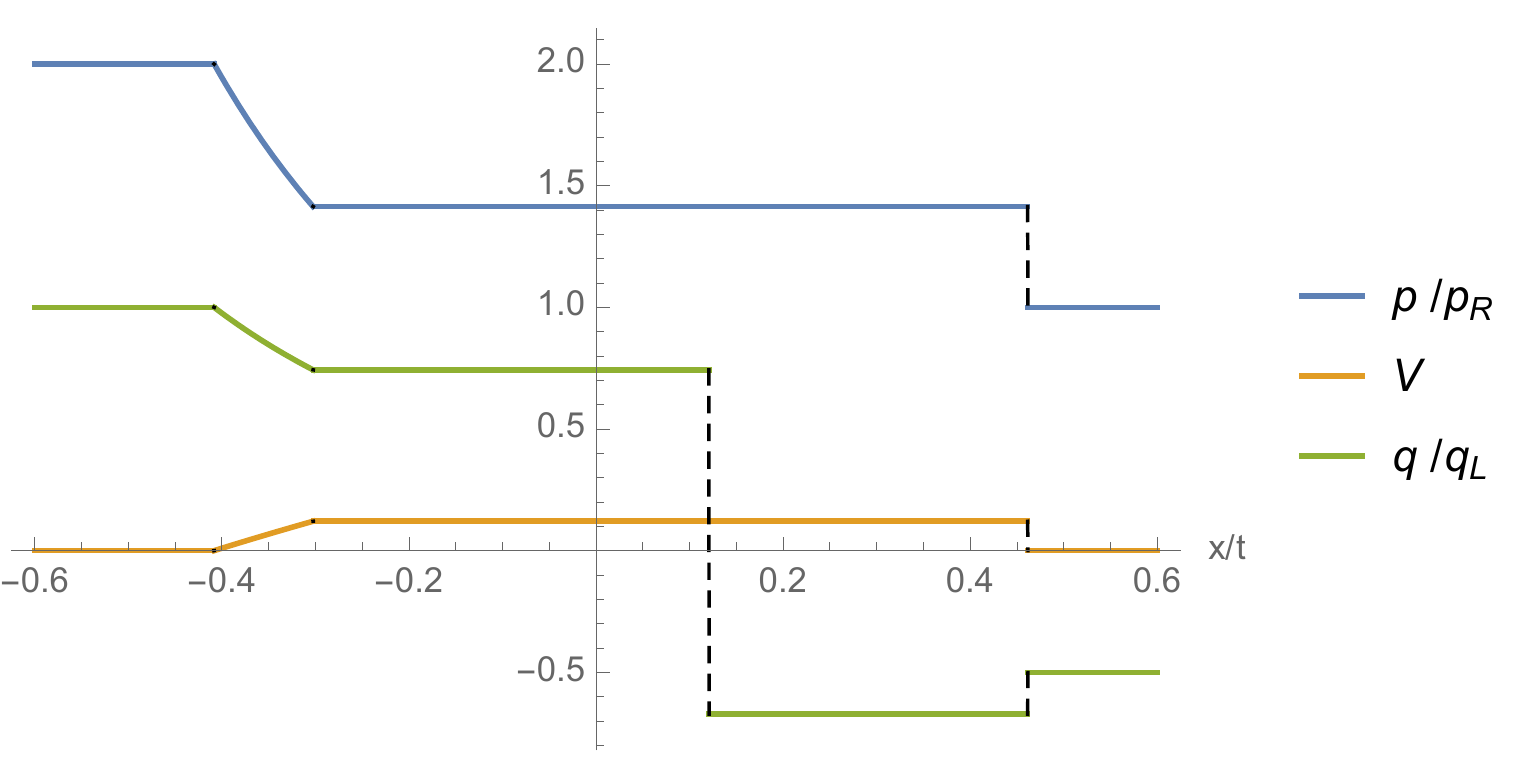}
 \caption{
 %\small 
 Plot of velocity, charge, and pressure profiles:  left $c_s^2 = 1/2$, right $c_s^2 = 1/6$.}

\end{figure}

\subsection{Simple Limits}

While we only have a parametric solution in general, various limits take a simpler form.
 Consider the first the limit $c_s \to 1$.  As $c_s^2 = 1/d$ for a conformal field theory, we can
 think about this limit as perturbing the number of spatial dimensions away from one, $d=1+\epsilon$.
 In this case, the width of the rarefaction fan scales as
 \be
 \delta \xi = \frac{1}{2} (\chi -1) \epsilon + O(\epsilon^2) \ ,
 \ee
 which vanishes as $\epsilon \to 0$.  
 Indeed, the difference between the rarefaction solution and the two-shock solution is controlled by $\epsilon$, with the differences
 \begin{align}
 \delta v &= 2 f(\chi) \epsilon^2 + O(\epsilon^3) \ , \\
 \delta u_R &=  f(\chi) \epsilon^3 + O(\epsilon^4) \ , \\  
 \delta p &= \frac{p_L}{\chi^2 (1+\chi)^4} f(\chi) \epsilon^2 + O(\epsilon^3) \ ,
 \end{align}
 where we have defined the function of $\chi$, positive for $\chi>1$,
 \[
 f(\chi) \equiv \frac{\chi}{8(1+\chi)^3} \left( 1- \chi + (1+\chi) \tanh^{-1} \frac{\chi-1}{\chi+1} \right) \ .
 \]
The quantities  $\delta v$, $\delta p$ and $\delta u_R$ are the differences between the rarefaction solution and the two shock solution in velocity, pressure and shock speed respectively, e.g.\ $\delta v = v(\mbox{2-shock}) - v(\mbox{rarefaction})$. In this limit, the two-shock solution slightly overestimates the pressure and fluid velocity in the NESS, and also the right moving shock speed.
 
The non-relativistic limit $\chi \to 1$ of the rarefaction solution also approaches the two-shock solution.
Let 
%$p_L-p_R\ll p_R+p_L$ and 
 $p_L = p_R(1+\epsilon )$, $\epsilon\ll1$, in which case
\begin{align}
\delta \xi &= \frac{(1-c_s^2) c_s}{2(1+c_s^2)}\epsilon + O(\epsilon^2),\\
\delta v &= \frac{(1-c_s^2)^2 c_s}{384(1+c_s^2)^3}\epsilon^3 + O(\epsilon^4),\\
\delta p &= p_L \frac{(1-c_s^2)^2}{384(1+c_s^2)^2}\epsilon^3 + O(\epsilon^4),\\
\delta u_R &=\frac{(1-c_s^2)^3 c_s}{768(1+c_s^2)^3} \epsilon^3 + O(\epsilon^4) ,
\end{align}
where $\delta \xi$, $\delta v$, $\delta p$ and $\delta u_R$ are as before.
In this limit, the two-shock solution again slightly overestimates the velocity and pressure in the NESS and the speed of the right moving shock.

While the two shock solution and the solution with a rarefaction region quickly approach each other in the limit $p_L \to p_R$, in 
the opposite limit where $\chi \gg 1$, the solutions have qualitatively different behavior.  In both cases, the speed of the right moving shock approaches one, $u_s$, $u_R \to 1$, but at a different rate.  Let us assume a large $\chi$ ansatz where $u_s = 1 - \delta$.  In the rarefaction case
\begin{align}
\chi &= \left( \frac{1}{\delta} \right)^{\frac{c_s^2+1}{4 c_s}+\frac{1}{2} } 
\frac{1}{\sqrt{2}} \sqrt{\frac{1}{c_s^2} - c_s^2} \left(-2 + \frac{4}{1+c_s^2} \right)^{\frac{1+c_s^2}{4c_s}} \left( 1 + O(\delta) \right) \ ,
\\
p &= p_L \delta^{\frac{c_s^2+1}{2 c_s}} \left( -2 + \frac{4}{1 + c_s^2} \right)^{- \frac{1+c_s^2}{2c_s}} \left( 1 + O(\delta) \right)\ ,
\\
v &= 1 - \frac{1+c_s^2}{1-c_s^2} \delta + O(\delta^2) \ .
\end{align}
In contrast, for the two shock solution $\chi \sim \frac{1}{\delta}$ and $p \sim \sqrt{\delta}$.

Another important difference is that the size of the rarefaction region grows in this limit $\chi \gg 1$.  The rightmost characteristic of the rarefaction fan approaches the location of the right moving shock:
\begin{align}
\xi_R = 1 - \frac{1+c_s^2}{(1 - c_s)^2} \delta + O(\delta^2) \ .
\end{align}
Thus the size of the NESS is correspondingly reduced.  Indeed, for all practical purposes, the NESS probably disappears in this limit.  An initial condition that is a step function is an idealization, and slightly smoothing the step function will destroy the NESS, as will viscous corrections, which smear out the shock.

\section{Non-linear Equation of State}

We can also consider the Riemman problem for nonlinear equations of state.  For simplicity we will assume that $v_T = 0$.  We consider a perturbation to the CFT by a relevant operator as in \cite{Pourhasan,Buchel},
\begin{align}
S_{QFT} = S_{CFT}+\lambda^\prime \int d^{d+1}x\,  \mathcal{O}(x)
\end{align}
in the limit $\lambda/T^{d+1-\Delta} \ll1$.
For relevance and unitarity, we require $\frac{d-1}{2} \leq \Delta < d+1$.  
Such a perturbation should affect the equilibrium pressure and energy density at second order in $\lambda'$.  Following
ref.\ \cite{Buchel}, we assume an ansatz where
\begin{align}
p &= c  \, T^{d+1} \left( 1 -   \frac{\lambda^2 }{T^{2(d+1-\Delta)}} \right) \ , \\
e &= c \, d \, T^{d+1} \left( 1 - \alpha  \frac{\lambda^2 }{T^{2(d+1-\Delta)}} \right) \ ,
\end{align}
where $\lambda$ and $\lambda'$ are proportional.\footnote{%
 We expect that a  relevant perturbation should decrease the effective number of degrees of freedom of the theory
 and thus further decrease the entropy at low temperatures, explaining the minus sign in front of the correction to the pressure. 
}
The Gibbs-Duhem relation $e + p = sT$ along with $s = \frac{dp}{dT}$ then imply that $\alpha = \frac{1}{d}(2 \Delta -d-2 )$.
Eliminating $T$, we can write an equation of state for $e$ as a function of $p$:
\begin{align}
e = p d + \epsilon \lambda^2 p^n \ ,
\end{align}
where 
\begin{align}
\epsilon &= 2 c^{\frac{2(d+1-\Delta)}{d+1}} (d+1 - \Delta) \ , \\
n &= \frac{2\Delta}{d+1} - 1 \ .
\end{align}
Note that $\epsilon > 0$.

As in the case of a linear equation of state, we find an adiabatic solution and match it onto a NESS region and a shock.  The equations of motion remain (\ref{con1}) and (\ref{con2}).
We can solve these equations in a perturbative expansion in $\lambda$ as follows:
\begin{align}
p = p_0 + p_1 \lambda^2, \quad
v_x = v_0+v_1 \lambda^2.
\end{align}
Again there are two solutions 
\begin{align}
p_0^\pm &= c_{0p}\left(\frac{1-\xi}{1+\xi}\right)^{\pm\frac{1+c_s^2}{2 c_s}}, \\
v_0^\pm(\xi) &= \frac{c_s\pm \xi}{\xi c_s \pm1}, \\
p_1^\pm &= -\epsilon\frac{c_s^2(2-2c_s^2+n(1+c_s^2)(n+nc_s^2-2))c_{0p}^n}{2(n-1)(c_s^4-1)}\left(\frac{1-\xi}{1+\xi}\right)^{\pm n\frac{1+c_s^2}{2c_s}}, \\
v_1^\pm&= \pm\epsilon \frac{n(\xi^2-1)c_s^3 c_{0p}^{n-1}}{2(\xi c_s \pm 1)^2}\left(\frac{1-\xi}{1+\xi}\right)^{\pm(n-1)\frac{1+c_s^2}{2c_s}}.
\end{align}
We are first interested in how fast the disturbance moves to the left ($\xi_L$).  This can be found perturbatively in $\lambda$ by requiring that $v_x(\xi_L)=0$ and  $p(\xi_L)=p_L$.  The result is
\begin{align}
\xi_L &= -c_s + \frac{n p_L^{n-1}c_s^3}{2}\epsilon \lambda^2 \ , \\
c_{0p} &=  p_L\left(\frac{1+c_s}{1-c_s}\right)^{-(1+c_s^2)/2 c_s}+\epsilon\lambda^2 p_L^n\frac{c_s^2(nc_s^2+n-2)}{2(n-1)(1+c_s^2)}\left(\frac{1+c_s}{1-c_s}\right)^{-(1+c_s^2)/2c_s} \ .
\end{align}

\begin{figure}[!htb]
\centering
  \centering
  \includegraphics[width=0.48\linewidth]{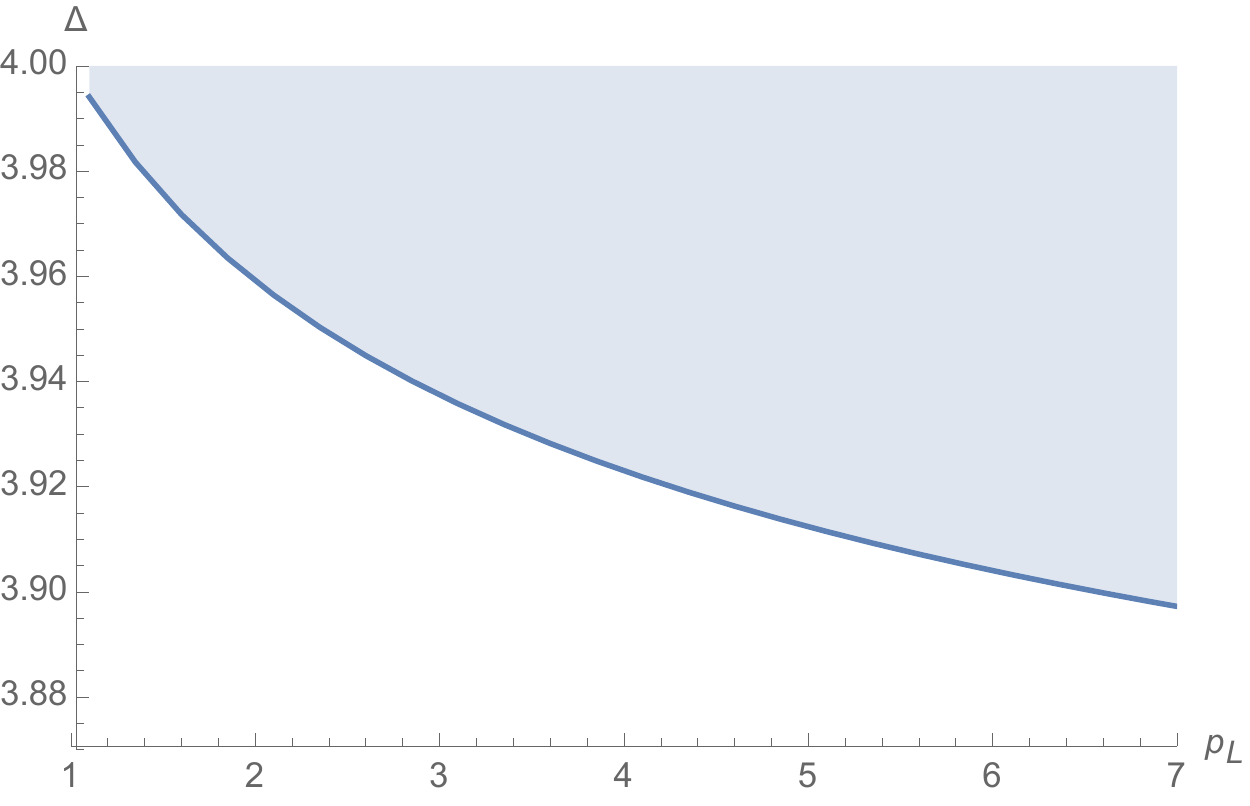}
  \quad
  \includegraphics[width=0.48\linewidth]{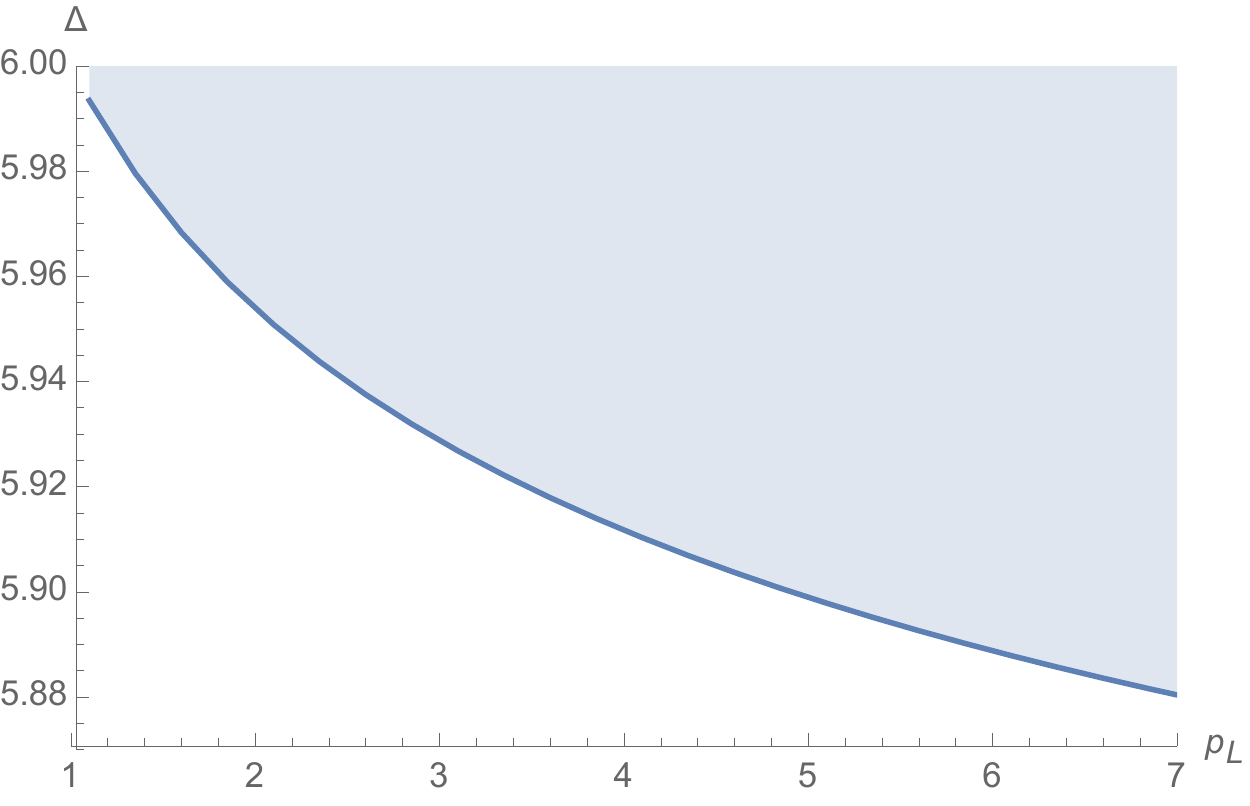} 
 \caption{
 %\small 
 Phase diagrams for the change in $P_{NESS}$ where the shaded region has $P_{NESS}<P_{CFT}$. The left is for $d= 3$ and the right $d =5$.}
\label{phase}
\end{figure}

Unfortunately, analytic solutions for the correction to the NESS pressure are not available.  However, one can still do the calculation numerically with an interesting result.  Unlike for the double shock solution presented in \cite{Pourhasan} the change to the NESS pressure is dependent on both $\Delta$ and $p_L$ ($p_L>p_R=1$) as opposed to only $\Delta$ in \cite{Pourhasan}.  We do not have an analytic expression for the curve, but the phase diagrams are presented in Figure (\ref{phase}) for two different spacetime dimensions.

In the two shock solution we can find the corrections analytically.  We find

\begin{align}
\delta v&=   \frac{(d+1)  \left(\chi ^n+\chi \right) \left((d+\chi ) \chi ^n-\chi  (d \chi
   +1)\right)}{4 \sqrt{d} \chi  (d+\chi )^{3/2} (d \chi +1)^{3/2}}\epsilon \lambda^2, \\
   \delta p&=\frac{ \left(\chi ^n-\chi \right) \left((d+\chi ) \chi ^n-\chi  (d \chi
   +1)\right)}{2 d (d+1) \chi  (\chi +1)}\epsilon \lambda^2.
\end{align}
The quantities  $\delta v$ and $\delta p$ are the differences between the QFT and the CFT values of velocity and pressure respectively, e.g.\ $\delta v = v_{QFT} - v_{CFT}$. We note that $\delta p \geq0$ for all relevant operators independent of $\chi$, where the inequality is saturated for $\Delta = d+1$.\footnote{This result is not at odds with \cite{Pourhasan} because they consider temperature rather than pressure.}

\section{Numerical Check}

We want to implement a numerical scheme to check our results.  To do this we use the same hydrodynamic scheme as \cite{herzog} which employed spectral methods. For simplicity we take $v_T = 0$.  We start with
\begin{eqnarray}
T^{\mu\nu} = (e + p) u^\mu u^\nu +p  \eta^{\mu\nu} + \Pi^{\mu\nu}
\end{eqnarray}
where we define $\Pi^{\mu\nu}$ recursively\footnote{%
The implicit definition of $\Pi^{\mu\nu}$ is Israel-Stewart like.  Formally, higher than second order gradient corrections are present in the definition of $\Pi^{\mu\nu}$.
}
 in a gradient expansion
\begin{eqnarray}
\Pi^{\mu\nu} &=& - \eta \sigma^{\mu\nu}- \tau_\Pi \left[ (D \Pi)^{\langle \mu\nu \rangle} + \frac{3}{2} \Pi^{\mu\nu} (\nabla \cdot u) \right] 
 - \frac{\lambda_2}{\eta} {\Pi^{\langle \mu}}_{\alpha} \Omega^{\nu \rangle \alpha}  \ 
\end{eqnarray}
where $D \equiv u^\mu \nabla_\mu$. 

Conformality implies tracelessness of $T^{\mu\nu}$ which in turn yields a relationship  $e = d \, p$ between the energy density $e$ and pressure $p$ in $d$ spatial dimensions.   
While $\mu$ in principle takes values from 0 to $d$, 
we let only $u^0$ and $u^1$ be nonzero.
 The vorticity is
\begin{eqnarray}
\Omega^{\mu\nu} \equiv \frac{1}{2} \Delta^{\mu \alpha} \Delta^{\nu \beta} (\nabla_\alpha u_\beta - \nabla_\beta u_\alpha) \ ,
\end{eqnarray}
where we have defined a projector onto a subspace orthogonal to the four velocity:
\begin{eqnarray}
\Delta^{\mu\nu} \equiv \eta^{\mu\nu} + u^\mu u^\nu \ .
\end{eqnarray}
The shear stress tensor is
\begin{eqnarray}
\sigma^{\mu\nu} \equiv 2 \nabla^{\langle \mu} u^{\nu \rangle} \ .
\end{eqnarray}
The angular brackets $\langle \rangle$ on the indices indicate projection onto traceless tensors orthogonal to the velocity
\begin{eqnarray}
A^{\langle \mu \nu \rangle} \equiv \frac{1}{2} \Delta^{\mu\alpha} \Delta^{\nu \beta}(A_{\alpha \beta} + A_{\beta \alpha}) - \frac{1}{d} \Delta^{\mu\nu} \Delta^{\alpha \beta} A_{\alpha \beta} \ .
\end{eqnarray}

Note that with these definitions, both $\Pi^{\mu\nu}$ and $\Omega^{\mu\nu}$ are traceless and orthogonal to the velocity
\[
u_\mu \Pi^{\mu\nu} = u_\mu \Omega^{\mu\nu} = 0 \ , \; \; \;
\Omega^\mu_\mu = \Pi^\mu_\mu = 0 \ .
\]
Using $v_T = 0$ and flows that only depend on $x$ and $t$, we know $\Pi^{xy} = 0$ and $\Omega^{xy} = 0.$  The remaining
one independent component of $\Pi^{\mu\nu}$ we choose to be 
$B = \Pi^{xx}-\Pi^{yy}$.  We then use the two conservation equations 
and the implicit definiton of $B$ to propagate forward in time.  

We give the equation of state in terms of the energy density $e$ and temperature $T$,
\begin{align}
e=\left(\frac{4\pi T}{3}\right)^{d+1} \ , 
\end{align}
and we start with the initial condition 
\begin{align}
u_x&= 0, \\
T &= \frac{T_R-T_L}{2}\tanh( \beta x) +\frac{T_R+T_L}{2}.
\end{align}

\begin{figure}[!htb]
\centering
  \centering
  \includegraphics[width=0.48\linewidth]{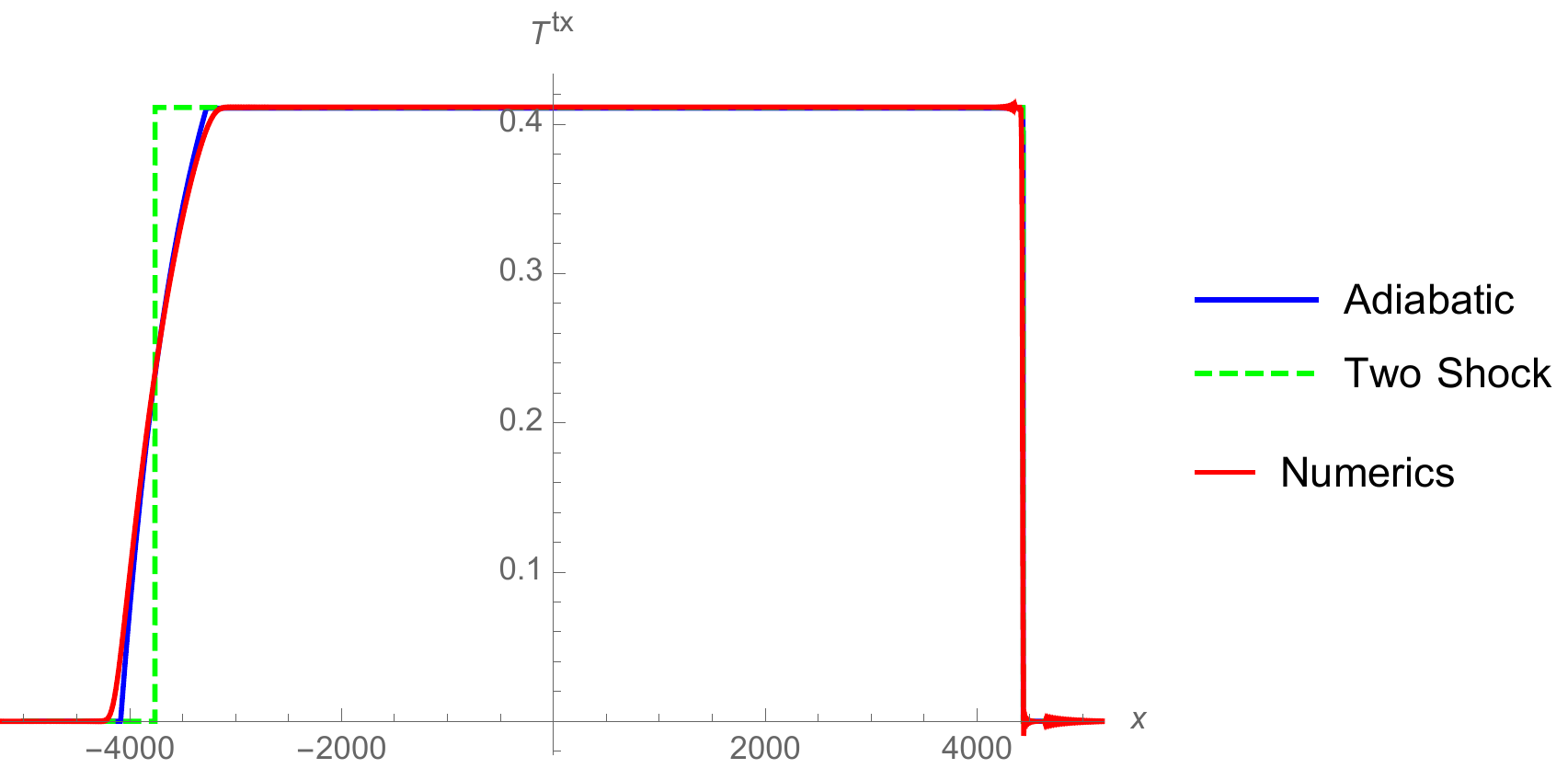}
  \quad
  \includegraphics[width=0.48\linewidth]{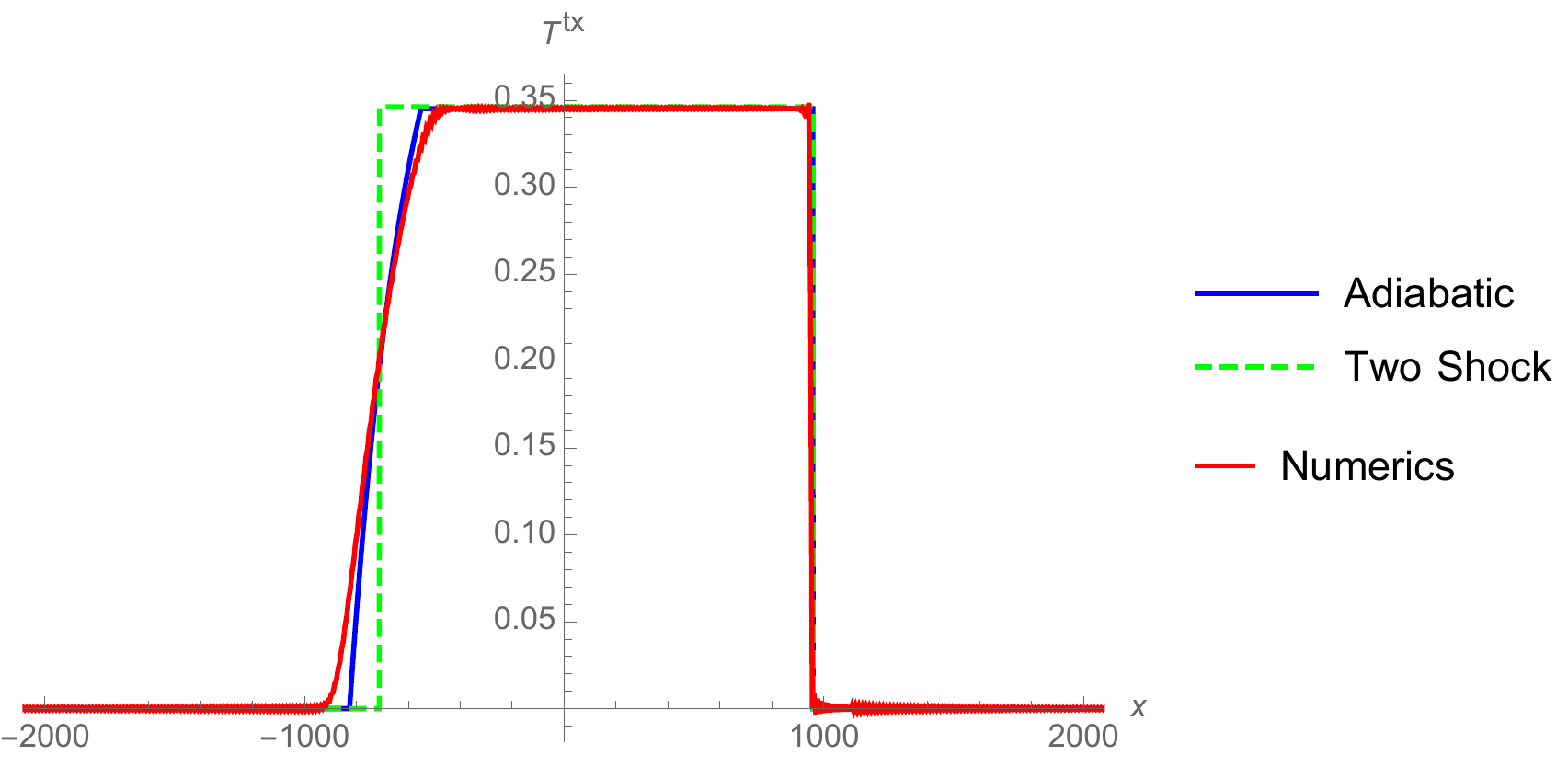} 
 \caption{
 %\small 
 Plot comparing $T^{tx}$ for the adiabatic solution and two shock solution to numerics.  The left uses $p_L = 2.75$, $p_R = 1$, $\beta = \pi/100$ with 3001 grid points at $t=5777$ (left) and $t=1652$ (right).}
\label{numerics}
\end{figure}

As can be seen in Figure (\ref{numerics}) our solution with an adiabatic region matches well with the numerics.  We accurately match the speed of the shock as well as the position of the adiabatic region.  The matching is not perfect in the adiabatic region because the initial profile was not a perfect step function.  The results were insensitive to the values of the dissipative coefficients $\eta, \tau_\Pi$ and $\lambda_2$.  While the difference cannot be seen in the figures, we also calculated $T^{tx}_{NESS}$ for the two analytic results and the numerics.  The results are presented in Table (\ref{table1})

\begin{table}[h]
\center
\begin{tabular}{l| c | c | c }
  $d$ &Two shock & Adiabatic & Numerics \\ \hline
  3 & 0.176587 & 0.176545 & 0.176551$\pm$ 0.000006 \\
 5&  0.14948 & 0.14936 & 0.149344$\pm $0.00009\\

\end{tabular}
\caption{The average and standard deviation of $T^{tx}$ over the flattest part of the NESS region compared with the theoretical values for $p_L = 1.75$, $p_R = 1$, $\beta = \pi/200$.  The adiabatic solution presented in this paper is a better fit than the two shock solution.}
\label{table1}
\end{table}

\section{Conclusion}

We presented a review of previous work on the Riemann problem for a linear equation of state.  The solution has the interesting feature of a steady state region with a momentum flux.  We contrast this solution with recent solutions \cite{Bhaseen2013, Pourhasan} that have appeared which incorrectly solved the Riemann problem by using a two shock ansatz.  These papers had failed to consider the entropy condition and missed the adiabatic expansion region (rarefaction region) which exists between the hotter reservoir and the NESS.  We also showed that for a conserved charge, the NESS region is actually two regions with different charges with a contact discontinuity separating them. 

In the nonrelativistic limit ($\chi \to 1$) and the small dimension limit ($c_s \to 1$), 
the two shock solution leads to small errors in the value for the momentum flux of the NESS as well as in the velocity of the shock wave.  We were able to show using a numerical simulation that the solution with the adiabatic region is preferred over the two shock solution. The adiabatic region was also a better match for the fluid profile than the shock propagating toward the high temperature reservoir. 

Finally, we considered a CFT which had been perturbed by a relevant operator.  The perturbation leads to a non-linear correction to the equation of state.  Again we found the adiabatic solution which should be matched onto the NESS.  Our solution gave a phase diagram for the correction to the NESS temperature that depended on $p_L$ and $\Delta$ the operator dimension.  This diagram contrasted with the two shock solution where the only relevant parameter was the operator dimension. 

One area for future investigation is adding viscosity to the solution.  In general adding viscosity makes analytic solutions impossible.  However, numerical simulations seem to show that the solution is very weakly affected by viscosity which gives hope that such a solution could be found.  
We are also interested in considering analytical solutions with smooth initial conditions to be better able to compare with our numerical results where the initial conditions are smooth by necessity (because we used spectral methods).  

Perhaps the most interesting question is a rephrasing of the problem using the fluid-gravity correspondence \cite{Hubeny:2011hd}
as a question about black-hole dynamics.
In this context, the Riemann problem considered here maps to a solution to Einstein's equations in an asymptotically anti-de Sitter space-time.  The temperature of the fluid can be re-interpreted as the location of an apparent horizon of a black hole.  Is there a gravitational counter-part of the entropy condition that we considered in this paper?  What is the gravity dual of the adiabatic region, and why is it required for consistency of the theory?

\vskip 0.1in
\noindent
{\bf Note Added:} While we were putting the finishing touches on this paper, we became aware of related and overlapping work
\cite{Schalm}.
% where the authors address and discuss the rarefaction region missing from their earlier paper \cite{Bhaseen2013}.

\section*{Acknowledgments}
We thank Koushik Balasubramanian for discussion during the early phases of this project as well as for his contributions to the coding of the numerical simulation.  We thank Amos Yarom for collaboration during the early stages of this project.  We would also like to thank Derek Teaney for discussion.
This work was supported in part by the National Science Foundation under Grant No.\ PHY13-16617.

\end{document}